# SKA Memo 87

# Aperture Arrays for the SKA: Dense or Sparse ?

Robert Braun & Wim van Cappellen
ASTRON

## Abstract

We briefly consider some design aspects of aperture arrays for use in radio astronomy, particularly contrasting the performance of dense and sparse aperture arrays. Recent insights have emerged in the final design phase of LOFAR which suggest that sparse aperture arrays have the best prospects for cost-effective performance at radio frequencies below about 500 MHz; exceeding those of both dense aperture arrays and parabolic reflectors by an order of magnitude. Very attractive performance, of 10,000 – 20,000 $m^2/K$, can be achieved with a sparse design that covers the 70 – 700 MHz range with two antenna systems that share receiver resources. Cost-effective systems of this type represent only a modest increment in system complexity over that being deployed in LOFAR and are achievable with today's technology.

## 1. Introduction

Over the past few years, the realization has emerged within the international SKA community that the most cost-effective and versatile forms of antenna elements for radio astronomy are very likely to be modest-size parabolic reflectors at the highest frequencies and some form of aperture array at the lowest frequencies. Quite some discussion has ensued over what the likely cross-over frequency might be between one technology and the other. Undoubtedly, the answer to this question will evolve over time as mass production, integration and digital processing all become more affordable. The great virtue of an aperture array at low radio frequencies is that a single (dual-polarization) antenna and receiver chain can deliver $\lambda^2/2$ of effective aperture, and when $\lambda$ is large this is a lot of aperture per active element. At low radio frequencies there is no way that a paraboloid of any size can compete with this in terms of raw performance. At high radio frequencies the opposite is true. The aerial density of active elements of an aperture array becomes so high that no one (but perhaps the military) could contemplate building more than a few square meters of such an antenna. At intermediate frequencies, there are various "secondary" issues (like instantaneous field-of-view (FOV) and geometric fore-shortening of an aperture array station) which might influence one's choice, depending on the scientific goal being considered. We'll leave that discussion for the time being and first concentrate on some basic aspects of aperture array design. Specifically, we will consider some specific cost and performance issues and their dependence on the adopted antenna spacing within an aperture array. Finally we contrast the likely cost and performance of some specific implementations of aperture arrays in the SKA context.

## 2. Aperture Array Design Issues

### 2.1 Relative cost of Antenna versus Receiver Chain

An important insight that has emerged from the LOFAR final design process is the relative cost of different components of an aperture array station. Each LOFAR remote station consists of a distribution of 96 low-band antennas (LBAs) covering the 20 – 80 MHz band and another distribution of 1536 high-band antennas (HBAs) covering the 120 – 250 MHz band. Each antenna element is attached to an LNA, the signal is transported to a first stage (4×4) analog beam-former for the HBAs, the signal is then digitized, frequency filtered and subjected to digital beam formation before being transported to a central site for correlation with other stations. Only one or the other of the antenna distributions is used at one time, since they can then feed the same receiver, beam-forming and signal transport chain. Eight independently steerable beams can be formed of 4 MHz each. With this configuration, the LBA mechanical components contribute less than 1% of the station cost, while the HBA mechanical components will contribute less than 13% given current estimates. For larger station bandwidths, the relative costs of antennas become less almost linearly. For example with a ten times larger (beam × bandwidth) product, of 320 MHz, the combined mechanical antenna costs including reflective ground screens have declined to about the 1% level.

The lessons which emerge from these practical considerations are: (1) that it is critical to maximize the performance of every receiver chain, since the number of receiver chains dominates the system cost in the currently explored regime and (2) that multiple antenna systems can feed the same receiver chains with only a modest incremental cost. We will see below, that the optimal antenna/receiver cost distribution for maximizing the system survey speed is more nearly equal, rather than being dominated by the receivers.

### 2.2 Relative cost of field-of-view

One of the most attractive aspects of aperture arrays is their ability to provide an instantaneous field-of-view (FOV) which is only limited by the digital signal processing which can be provided. Once any (optional) analog beam-forming has been done and the signal is digitized and frequency filtered it is passed on to digital beam-forming hardware to form a series of station beams. The computational capacity of the digital beam-former depends on whether the output beams are required to be independently steerable or whether they instead form a regular, fully-sampled grid on the sky, in which case an FFT-like beam-forming approach can be adopted. For independent beams the complexity scales as N×M, the product of the number of (compound-)antennas, N, and the number of output beams, M, while for a fully-sampled grid of beams from a regularly sampled array, be it sparse or dense, it scales as $Nlog_2(N)$. In the sparse case, the output beams contain grating responses at regular intervals, which could be used to feed multiple correlator channels each employing a suitable distinct delay and phase tracking center. For an irregularly sampled sparse array the digital beam former complexity scales as $Nf^2 \, log_2(Nf^2)$ for an array of sparseness factor, f. The sparseness factor is the mean ratio of actual antenna spacing relative to $\lambda/2$. From this comparison it is clear that for even modest numbers of output beams, the FFT-like approach is likely to be the most cost-effective. It is also clear that the cost penalty invoked in an FFT beam-former by adopting a sparseness factor greater than one, while demanding the same FOV, scales roughly as $f^2$.

We can illustrate these considerations best with a specific example. The APERTIF project (van Cappellen, 2006) to outfit each of the 25m paraboloids of the Westerbork array with a multi-beam focal plane array (FPA) receiver is currently envisioning a 64 element dual-polarization antenna system to provide 25 completely independent beams, each of 300 MHz bandwidth. For this application, the digital output beams must be fully independent since the beam-former weightings will be fine-tuned for each (off-axis) beam in the telescope focal plane. The digital beam-forming hardware, with complexity N×M = 1600, constitutes 15% of the total cost of each FPA system (which is analogous to a complete aperture array station in the current context). This is the same digital beam-former complexity as an FFT-like beam-former for an N = 96 antenna aperture array with an average sparseness f = 1.5, for which $Nf^2 log_2(Nf^2)$ = 1660. The point of this example, is to demonstrate that full FOV utilization can be achieved in sparse aperture arrays without an excessive cost in beam-forming.

## 2.3 Relative Bandwidth and Sensitivity

An important design parameter of an aperture array is the frequency range over which it is to be employed. There are various types of antenna elements which can provide good power matching to an LNA over an intrinsically broad band. These are typically characterized by an inherently logarithmic design; for example the Log Periodic Dipole (LPD) antenna, the Bunny Ear Antenna (BEA) or variants of a quad-ridged horn. Different portions of the structure, which are roughly matched to a half-wavelength, become active at different frequencies. A distinct ground-plane or, in the case of the LPD, another part of the antenna itself, provides the basic directivity in a forward direction. Hemi-spheric directivity is critical for (ground-based) radio astronomy at frequencies greater than about 100 MHz (where we are not sky noise dominated) since otherwise we would inevitably pick up at least 50% of the 300 K below our feet.

When antennas of this type are separated by at least $\lambda_{max}/\sqrt{2}$, they can provide up to $\lambda^2/2$ of effective aperture with an opening angle of the forward response that is up to 120 degrees at half power over as much as a decade of frequency. The $\lambda_{max}/\sqrt{2}$ condition is what defines a "sparse" aperture array. By contrast, a "dense" aperture array is one in which the element spacing is of order $\lambda_{min}/2$. In this dense configuration, there is no increase in the effective aperture at lower frequencies (except for the finite number of edge elements of a distribution).

Of course if a sparse aperture array is being employed as a radio telescope "station" there are some additional considerations. When the total frequency range is large, the surface-filling factor might become so small at high frequencies that the brightness sensitivity of the resulting station may be too small. A practical lower limit to the surface-filling factor might be 10%. This implies a "natural" frequency range for a sparse aperture array station of about 3.5:1. Such a band-width is achievable with many types of electrically large antenna elements.

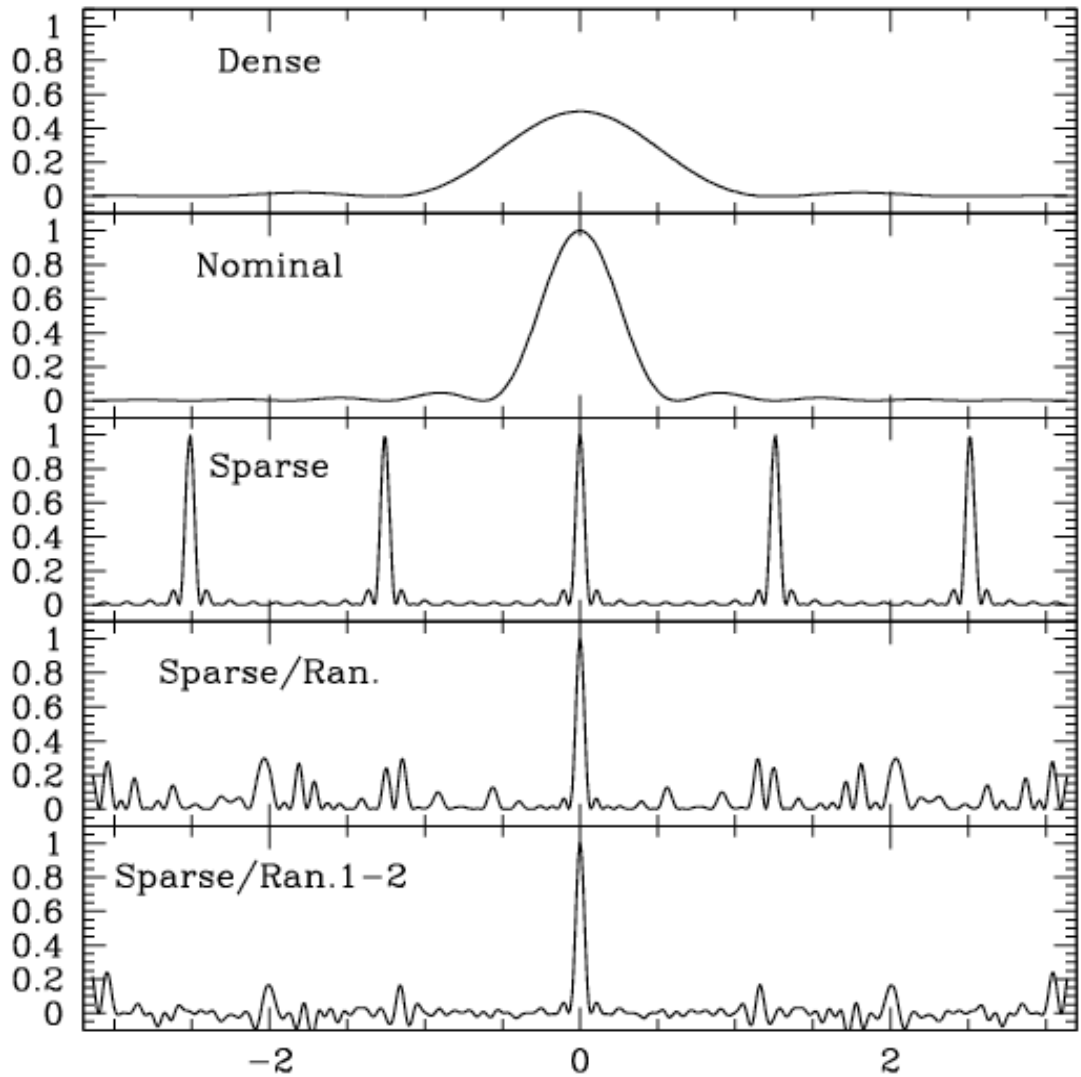

**Figure 1.** Illustration of station beam patterns for various forms of aperture sampling. Beam patterns are normalized to the peak sensitivity of a nominally sampled regular array. From top to bottom are: densely sampled, nominally sampled, sparse regular sampling, sparse random sampling and finally the product of two different random sparse sampled beams.

**2.4 Grating Responses and Scan-angle Resonances**

Other considerations for a sparse aperture array station are those of potential grating responses, blind scan angles, frequency resonances or other undesirable attributes of the station beam properties. All of these issues are being confronted and addressed in the context of LOFAR. The LOFAR design choice was for a semi-sparse aperture array, where the $\lambda/2$ frequency has been chosen near the low frequency end of the antenna bandwidth, about 50 MHz for the low-band antennas (20 – 80 MHz) and about 130 MHz for the high-band (120 – 250 MHz). The reason for this choice can be expressed in one word, namely, *sensitivity*. As outlined above, the station cost is dominated by that of the receiver chains, so the sensitivity per receiver chain should be maximized. The choice of a sparse array was made despite the complication of grating responses which can potentially arise from such a station configuration. This is at odds with the accepted wisdom of antenna engineers who generally view grating responses as the worst of all possible evils.

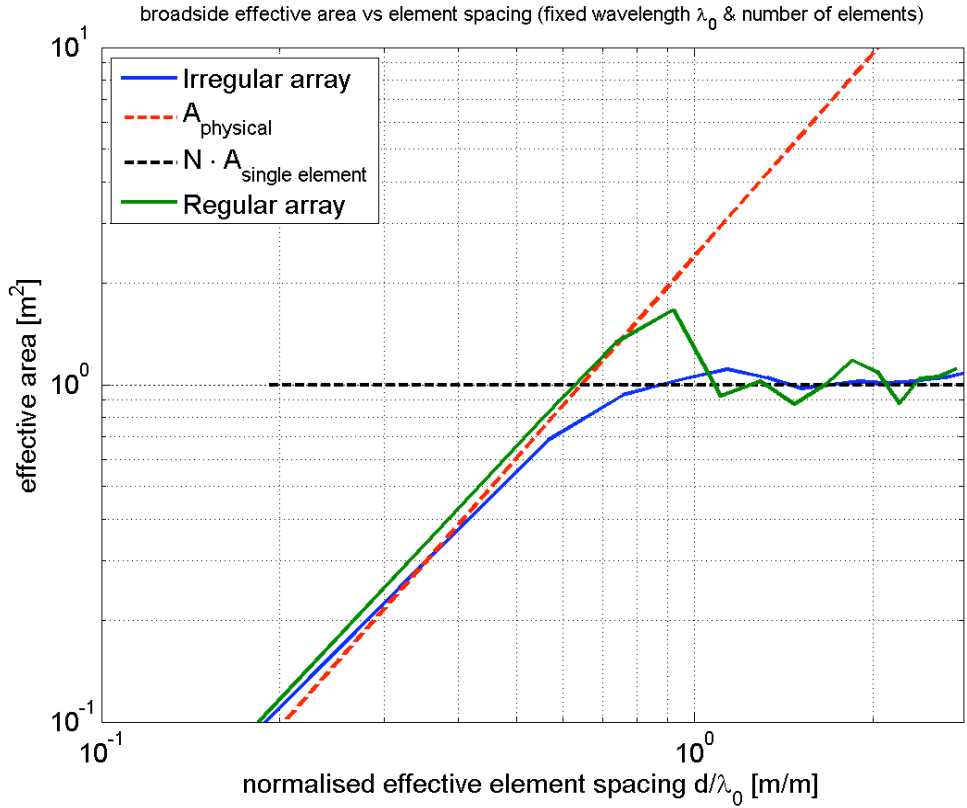

**Figure 2.** The broad-side effective area of variously sampled aperture arrays as function of the mean inter-element spacing in $\lambda$'s (from Van Cappellen et al. 2006). In the dense regime (d < $\lambda$/2), $A_{eff}$ ~ $A_{phys}$. In the sparse regime (d > 2$\lambda$), $A_{eff}$ ~ (N × $\lambda^2$/2). Irregular arrays show a smooth transition from the dense to sparse regime, while regular arrays show strong oscillations.

Some of the relevant issues are illustrated in Figure 1, where we contrast in a schematic way the station beam properties of dense, nominal and various sparsely sampled apertures. A densely sampled aperture has a maximum effective aperture of the geometric area, (N × $\lambda_{min}^2$/4) and as such has lower than nominal sensitivity together with a wider than nominal beam. A nominally sampled aperture (spacing of $\lambda_{nom}$/√2) has a nominal effective aperture (N × $\lambda_{nom}^2$/2) together with a nominal beam size. A regular, sparsely sampled aperture retains the nominal sensitivity, but has a useful beam size that has decreased by the degree of sparseness; a factor of five in the toy example shown in the figure. Additional grating responses also become apparent, such that the integral over all grating responses is still identical with that of the critically sampled case just discussed. If randomization of the antenna positions is employed, but the degree of sparseness is preserved, then the main station beam retains the nominal sensitivity and the beam size that is coupled to the degree of sparseness. However, the grating responses are redistributed into side-lobes. In the event that we are considering the auto-correlation of a single station or the cross-correlation of two strictly identical stations, then the amount of power in these unwanted grating- or side-lobes is preserved and is determined directly by the degree of sparseness. The bottom panel of Figure 1 illustrates the more interesting case in which the two stations being correlated to form power have significantly different side-lobe patterns (such as might be realized in practice by introducing a rotation of one sparse station relative to another). In this case there is a real increase in the main-lobe to side-lobe power ratio. With optimum side-lobe cancellation, there can be very substantial improvements in this ratio.

It should be noted that the grating and side-lobe issues that we illustrate schematically in Figure 1 are based on toy analytic simulations involving only eight antennas. Realistic aperture stations, containing thousands of antennas will have much lower side-lobe levels, as we will discuss in the context of calibration concerns below.

Many of the effects we have just considered have been demonstrated with full EM simulations by Van Cappellen et al. (2006). Since these are so relevant to the current discussion we reproduce some of those results here, although we refer the reader to the original reference for more details. In Fig. 2 we illustrate the broad-side effective area of a 96 dual-polarization antenna array for various types of aperture sampling as a function of the mean inter-element separation. The plot nicely shows the transition from the dense to the sparse regime. For element spacings of $0.2 - 0.5\ \lambda$, the effective area closely tracks the physical area of the station, while for spacings greater than about $1.5\ \lambda$, the effective area settles down to $(N \times \lambda^2/2)$. In the transition regime of $0.5 - 1.5\ \lambda$, the array geometry becomes very important. A regularly sampled array shows strong oscillations in broad-side performance in this regime. On the other hand, an irregularly sampled array shows a smooth turn-over in effective aperture in this transition from dense to sparse sampling. Although Figure 2 has been calculated for a sequence of inter-element spacings, the corresponding behavior is also seen when the simulation sequence is done over frequency with a fixed effective element spacing.

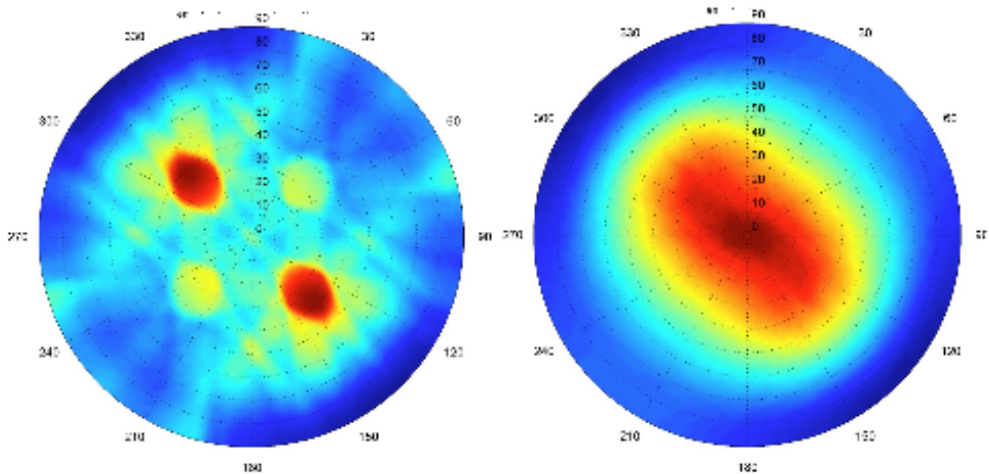

**Figure 3.** Effective area as a function of scan angle for a regular sparse array (left) and an irregular sparse array (right) (from Van Cappellen et al. 2006). The plot covers an entire hemisphere from the zenith to the horizon. Irregular sparse arrays show a smooth variation of performance with scan-angle in keeping with simple geometric fore-shortening of the station, while regular sparse arrays show severe resonances.

The same distinction can be seen in the scanning performance of sparse aperture arrays, as illustrated in Fig. 3. A regularly sampled sparse array has severe variations in performance with scanning direction, while an irregular sparse distribution does not.

These considerations have led to the design choice of the LOFAR low-band stations. A semi-random antenna distribution has been adopted for these stations which straddle the dense-sparse transition. This yields smooth performance variations with scan angle and frequency, while providing a nearly nominal sensitivity over much of the band. Even for these stations of only 100 antennas, grating responses are quite effectively redistributed into side-lobes, with an RMS power level for the auto-correlation response of about 1/N, or about -20 dB. A second method of station grating- and side-lobe suppression which will likely be employed is adoption of purposely different station layouts for different stations. Since it is the product of each pair of station voltage beams which determines the interferometer power pattern, a judicious choice of different station layouts can maximize the main lobe response and minimize side-lobe power. One simple way to do this is with a different rotation of some fixed layout pattern for each different station. Station rotation can contribute an additional factor of 10 to 20 dB in power side-lobe suppression of the cross-correlation response. The third method of grating- and side-lobe suppression that will be employed is the use of multi-frequency-synthesis (MFS). Up to this point we have been considering the instrumental response at a single discrete frequency. In practice we will be dealing with relative bandwidths of 10's of percent. In this regime, there is a further suppression of structures outside of the main-lobe by about 20 dB. Together these three strategies yield mean side-lobe power levels of about -50 dB. This is comparable to the level of error side-lobes which ensue from 1% calibration errors in the assumed complex gains of the antenna elements during beam formation. All three of these approaches are discussed in greater detail in a recent memo dealing with LOFAR calibration by Stefan Wijnholds (2006).

**2.5 Calibration Concerns**

The previous discussion of station beam imperfections is particularly relevant to the issue of (self-)calibration of data obtained with the SKA at frequencies below a few GHz. Reaching the thermal noise limit even after integrations lasting 10's of hours with of order 100 times the instantaneous sensitivity of current telescopes will imply routinely achieving a dynamic range for continuum applications of $10^6 - 10^7$. This will be quite a challenge. Important factors influencing the achievable dynamic range are the size, shape and temporal stability of the main- and side-lobes of the station beams. An important simplification for the calibration process is the effective isolation of a suitably small number of celestial sources that need to be considered simultaneously, together with a suitably small number of required parameters to describe the (time-dependent) shape of the station beam. For example, the simplest situation possible is the one where the station beam has no significant variation in time, encompasses only a single, phase coherent patch of the sky and has completely negligible side-lobes. This is essentially the situation we assume to apply at GHz frequencies with 25m class paraboloids, except that we are often already dynamic range limited at our current modest sensitivity due to pointing errors of the dishes and side-lobe variability caused (for example) by beam rotation on the sky. The mean side-lobe levels for current systems near 1 GHz are about -40 dB. Achieving the much higher dynamic range required for the SKA will probably require a larger station size (of perhaps 60 – 80m) to limit the main-lobe FOV as well as a lower mean side-lobe level of about -50 dB. The other requirement to meet these challenging dynamic range goals will be effective software that enables modeling of time-dependent phenomena with the smallest possible number of free parameters.

As just discussed above in §2.4, strategies have been developed to achieve these required mean side-lobe levels by employing sparse aperture array stations that use (1) randomized antenna geometries, (2) relative station rotation, as well as (3) multi-frequency synthesis. Although real-life experience with these systems must still be obtained, there are good prospects for realizing the necessary station beam performance.

It is important to consider that small paraboloids will have a very difficult time meeting a mean side-lobe specification of -50 dB at sub-GHz frequencies. The relatively small dish diameter (measured in λ's) together with likely diffractive blockages (of the receiver package and feed support legs) will probably yield mean (discrete frequency) side-lobe levels of about -25 dB. If stations are formed of small numbers (eg. 10's) of such dishes then these stations will have appalling beam characteristics, with ~10% side-lobes inside the envelope defined by the small dish primary beam pattern. This is likely to create an extremely challenging imaging and calibration problem. Whether it can be done at all must still be demonstrated.

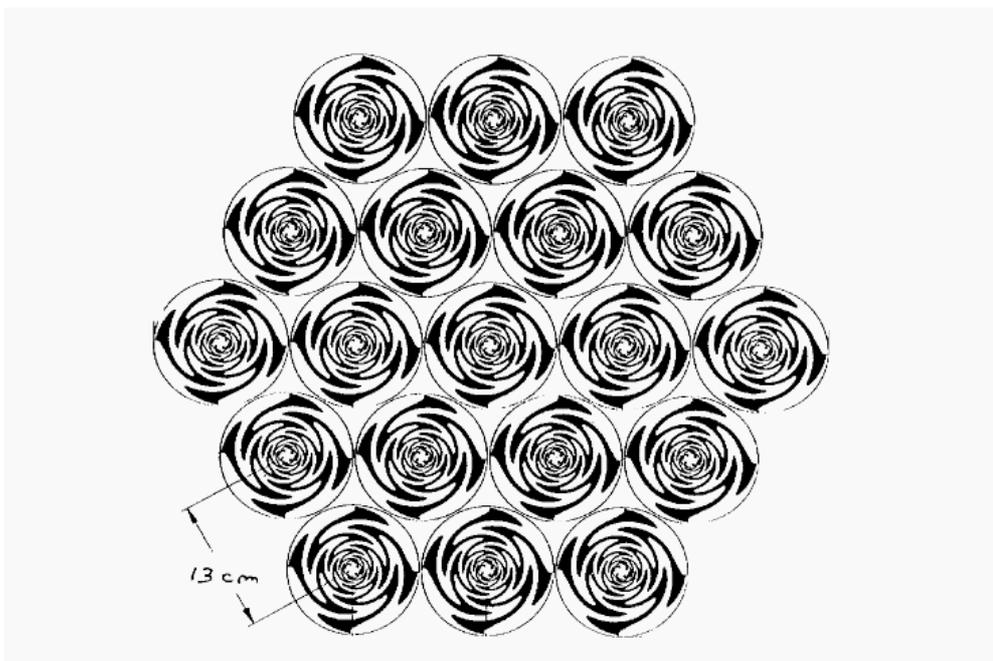

**Figure 4.** Example of one type of *planar* broad-band antenna elements. An hexagonally packed distribution of dual-polarization sinuous antennas is shown (from Fisher & Bradley 2000). Although these antennas are usually employed with a tuned reflective ground-plane, we instead consider their application at several λ above a diffusive ground-screen to provide broad-band hemi-spheric directivity from a flat radiating surface.

**2.6 2-D or 3-D Antennas**

An important cost consideration for aperture arrays, be they sparse or dense, is the intrinsically three-dimensional antenna structure that is required for the combination of wide bandwidth, dual polarization, power matching and hemi-spheric directivity. It would be quite unfortunate if at the low frequencies, where the performance per active element is high, there would be a prohibitively high antenna cost due to the large physical antenna volume. A possible solution to this dilemma might be the use of planar broad-band, dual-polarization antennas, such as planar LPDs or the sinuous spiral antenna, (as depicted in Figure 4) which are normally used in combination with a ground plane (for low noise applications). In fact, for military applications, such antennas covering a decade of band-width, are already employed using a broad-band absorber in a cavity behind the antenna. If such planar elements are instead suspended at several wavelengths above a diffusing ground plane, which is strongly textured over the wavelength range over which it is being used, they might provide *low-noise*, hemi-spheric directivity. Data transport might be accomplished via co-ax cable vertically down from each LNA to below the diffusing ground screen. Analog data transport over optical fiber might be even more attractive, once this is affordable.

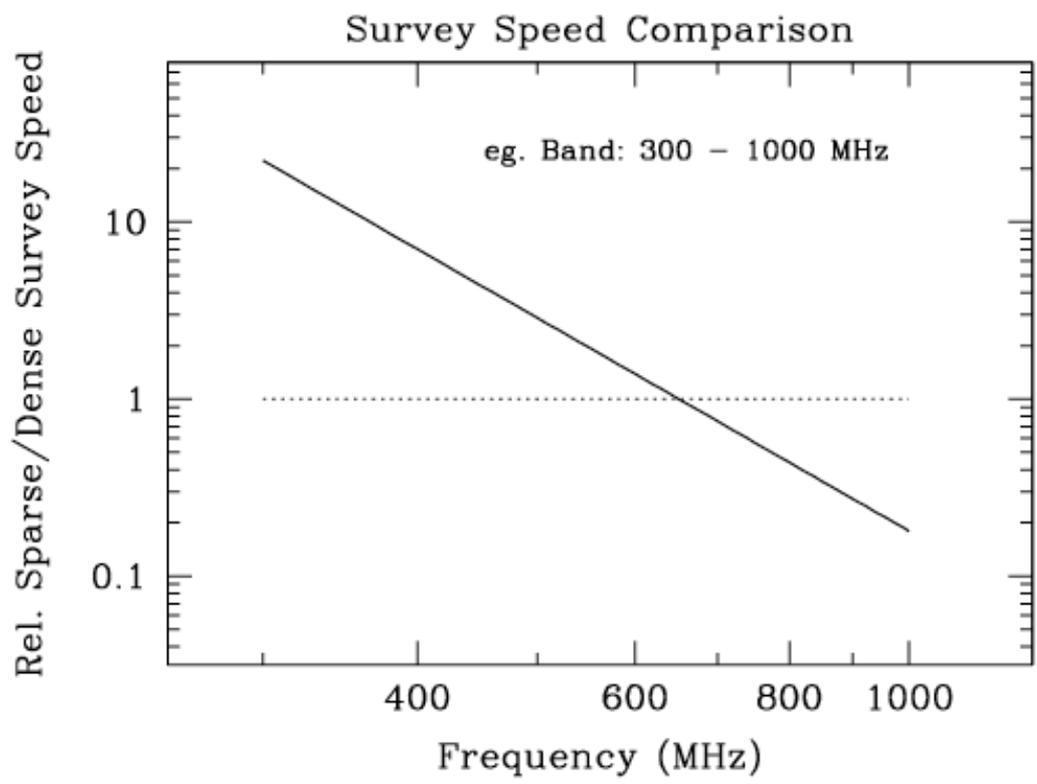

**Figure 5.** Survey speed comparison of sparse and dense aperture arrays. The ratio of FOV× Sens$^2$ is shown (solid line) for aperture arrays having the same number of antennas, receiver chains and $T_{sys}$ and a relative band-width of 3.3:1. As an example, the frequency range 300 – 1000 MHz is plotted, although the form is fully general. Dense arrays can be faster by up to a factor of 5 at the high frequency end of the band, while sparse arrays can be faster by up to a factor of 20 at the low end of the band. For both cases we assume the same data transport and correlator capacity. A ratio of unity is indicated by a dotted line for comparison.

## 3. Dense versus Sparse: FOV, Sensitivity and Survey Speed

As outlined above, a dense aperture array has an effective aperture that is fixed by the choice of the antenna spacing of $\lambda_{min}/2$, to be (N × $\lambda^2_{min}$/4), for some number of active elements, N, while a sparse array will have an effective aperture of (N × $\lambda^2$/2) over the entire frequency range of operation. If we assume, for example, a 3.3:1 frequency range, then the sparse array will provide 20 times the effective aperture at the lowest operating frequency compared to a dense array, for the same number of active elements. This higher sensitivity does not come without some penalty, since a larger station footprint would also imply a smaller instantaneous field-of-view (FOV) of each station beam by this same factor of 20, in the specific example considered above.

Since many of the anticipated scientific applications of the SKA involve some form of large area survey, it might be appropriate to consider the "Survey Speed" $\propto$ FOV× $S^2$ to combine these two aspects into a single figure of merit. The sensitivity of a sparse station of $N_A$ antennas is: $S_{sp} = N_A\lambda^2/(2T_{sys})$, while for a dense station it is $S_{de} = N_A\lambda_{min}^2/(4T_{sys})$. On the other hand, the *potential* field-of-view of a station with $N_R$ receiver chains (where in practice $N_R$ may be quite different than $N_A$) is FOV = $N_R\lambda^2/A_{phys}$, where $A_{phys}$ is the physical area of the station; $N_A\lambda_{max}^2/2$ for a sparse array and $N_A\lambda_{min}^2/4$ for a dense array. This yields a survey speed, $SS_{sp} = N_RN_A\lambda^6/(2T_{sys}^2\lambda_{max}^2)$ for the sparse array and $SS_{de} = N_RN_A\lambda_{min}^2\lambda^2/(4T_{sys}^2)$ for the dense. If we assume that each station has the same $N_A$, $N_R$ and $T_{sys}$, then we can form the relative survey speed for the two types of array: $SS_{sp}/SS_{de} = 2\lambda^4/(\lambda_{max}^2\lambda_{min}^2)$. We plot the relative survey speed over the band-width of the two arrays in Figure 5, where we have assumed a relative band-width of 3.3:1. In the highest frequency portion of the band there is a modest gain in survey speed, by up to a factor for 5, of the dense array, while over the lower portion of the band there is a dramatic gain in survey speed by up to a factor of 20 for the sparse array. This is a completely general result, which depends only on the relative band-width of the arrays under consideration. The *potential* FOV is realized in practice if the data rate of $N_R$ output station beams can actually be transported and correlated. As we will see below, this may not always be the case for actual aperture array implementations for the SKA. The actual survey speed contrast must then be assessed for specific systems using the specific data-rates which can be accommodated.

The linear dependence of survey speed on both $N_A$ and $N_R$ for both types of aperture arrays considered above, implies that an optimum cost distribution between the antenna and receiver aspects of a system, $C_A$ and $C_R$, is 50:50 if we include all station infrastructure costs in $C_A$ and the digital data transport and correlator costs in $C_R$ (cf. Bregman, 2004). If we compare this to the LOFAR system design, where receiver costs constitute about 68% of the total, we can conclude that the distribution is non-optimal. The use of a greater degree of analog beam formation (for example perhaps 4×4 in the low band and 6×6 in the high band) would have yielded more cost-effective system performance.

## 4. Dense versus Sparse: A Specific Comparison

The performance of various aperture array systems is best contrasted by considering some specific design choice. Consider the SKA specifications needed for imaging sub-structure at the epoch of re-ionization, detecting the earliest individual galaxies and conducting an effective "Dark Energy Survey"; a sensitivity of 10,000 – 20,000 $m^2$/K at 70 – 700 MHz (HI emission at z = 1 – 20) together with an instantaneous field of view of as much as 100 square degrees. Assuming, for example, $T_{sys} = T_{gal} + 40$ K, with $T_{gal} = 35(\nu/300 \text{ MHz})^{-2.6}$ K, implies a number of sparse array active elements of about $N_{Atot} = 3\times10^6$. Taking an element separation of $\lambda/\sqrt{2}$ at 250 MHz yields the sensitivity curve over frequency shown in Figure 6 as the solid line. The same receiver chains could be used to process a second set of $3\times10^6$ antennas which are sparse down to 70 MHz. The sensitivity of this antenna system is given by the dashed curve in Figure 6. If we distribute the antenna elements over a total of 300 stations, we would have $N_A = 10^4$ antenna elements per station. This would yield a station diameter of 96m in the high band (250 – 900 MHz) and 340m in the low band (70 – 250 MHz). Within a station one might form compound antennas via time-delay analog beam-forming from groups of 5×5 dual-polarization elements. A station would then consist of $N_R = 400$ such compound antennas and the accompanying digital receiver chains. The FOV of the compound antennas would vary over the frequency range of a station between about 300 deg$^2$ at the low frequency end to 30 deg$^2$ at the high end. A single analog FOV is therefore already marginally sufficient to meet the SKA design goal of 100 deg$^2$. If necessary, several analog FOV's could be generated, although this would have a linear impact on the total number of receiver chains, and hence the system cost. Tiling the entire compound element FOV would require 400 digital station beams (i.e. the same number as the number of compound antennas in a station). Digital FFT beam-former complexity will vary over the band as $N_R f^2 log_2(N_R f^2) = 3,500 – 48,000$. Assuming a band-width per station beam of 300 MHz, implies a station output band-width (per polarization) of 120 GHz per complete analog FOV.

A second, more cost-effective variant of this system might employ analog time-delay beam-forming from groups of 8×8 dual polarization elements. Each station would then require only 156 receiver chains and yield a compound antenna FOV that varies between 120 deg$^2$ at the low frequency end to 12 deg$^2$ at the high end. This degree of analog beam formation is likely to represent the 50:50 optimum cost distribution between antenna and receivers components that maximizes survey speed for a given total system cost, given current relative cost estimates from both LOFAR and APERTIF. The output station band-width is 47 GHz per compound antenna FOV.

If instead we consider a dense aperture array in the band 300 – 1000 MHz, with $\lambda_{min}/2$ at 1000 MHz and require a sensitivity of about 10,000 m$^2$/K at 1000 MHz with the same $T_{sys}$ dependence assumed above, then this requires some $20\times10^6$ antenna elements. The sensitivity of such a system as function of frequency is indicated by the dotted line in Figure 6. If such a system were distributed over 300 stations, then each station might have $N_A=67,000$ antennas and a diameter of 44m. If one were to apply analog beam-forming in groups of 8×8 dual-polarization elements, then this would result in some $N_R=1050$ compound antennas and accompanying digital receiver chains per station. The FOV of a compound antenna would vary from 200 deg$^2$ at the high frequency end to 2000 deg$^2$ at the low end, although it would require some 1050 digital station beams to tile the compound antenna FOV. Digital FFT beam-former complexity would be $N_R f^2 \, log_2 \, (N_R f^2) = 10,000$ across the band. The station output band-width (per polarization) per complete analog FOV is then 315 GHz (assuming 300 MHz per beam).

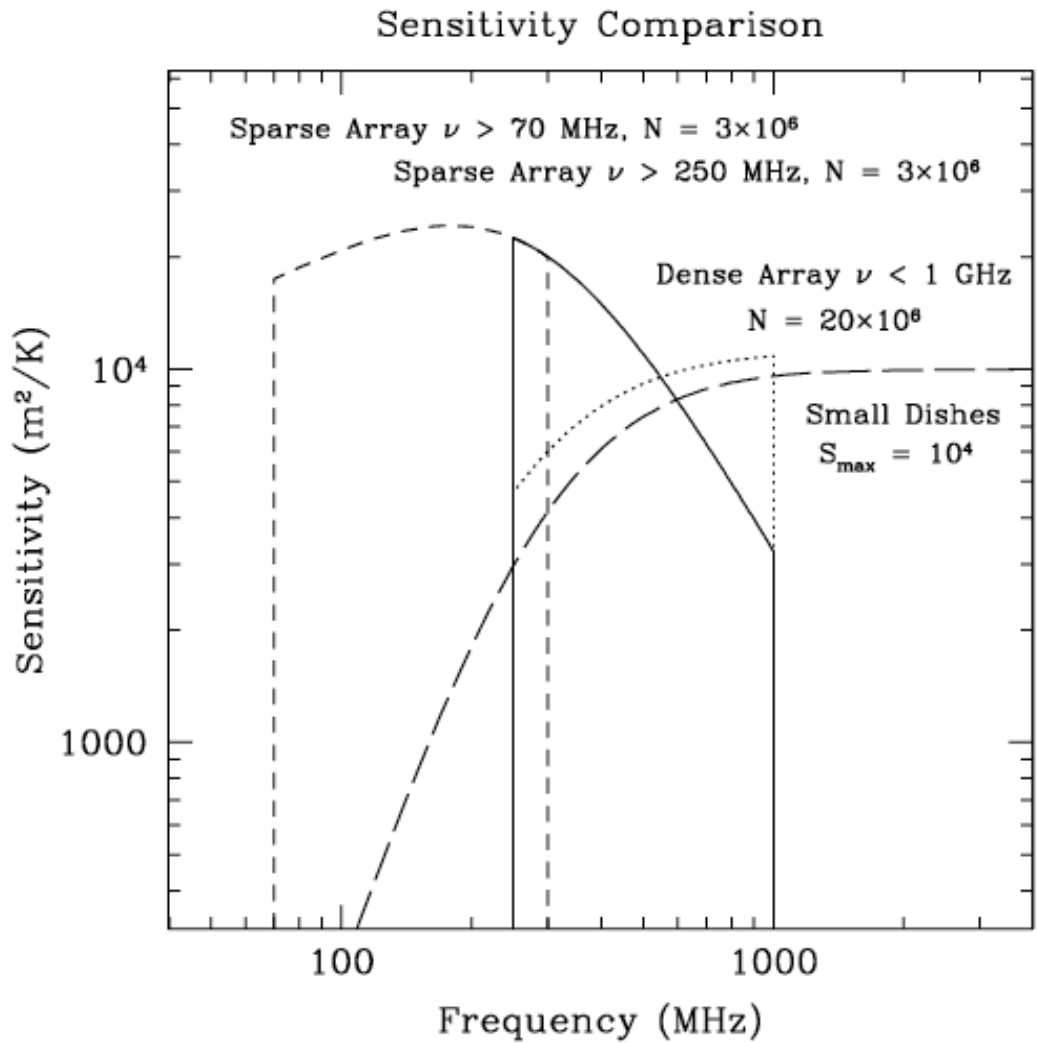

**Figure 6.** Sensitivity comparison of aperture arrays and small dishes. Two sparse aperture arrays are shown, each of $3\times10^6$ elements, one sparse above 250 MHz (solid line) and the other sparse above 70 MHz (dashed line) which could share the same digital receiver chains (that dominate the cost). Their performance is contrasted with a $20\times10^6$ element aperture array which is dense below 1 GHz (dotted line) and an array of small dishes with an assumed high frequency performance of $10^4$ m²/K (long-dashed line).

To bring down the system cost of the dense array by lowering the number of receiver chains to that of the cost-optimized sparse array considered above, $N_R\sim150$ for each station, would imply forming compound antennas of 21×21 elements. This would yield a FOV that varies between about 300 deg² at the low frequency end to 30 deg² at the high end and an output station band-width of 47 GHz per compound antenna FOV.

The relative cost of the two conceptual systems outlined above is expected to be roughly proportional to the number of digital receiver chains; 400 (or 156) per station in the sparse system and 1050 (or 152) per station in the dense array. The digital beam-former has a comparable complexity in both cases. Although at first glance one might expect the larger physical size of the sparse stations to result in a higher internal data transport cost this is likely to be more than offset by the fact that there are seven times the number of LNA's and analog signal paths in the dense array. While reliable estimates will require much more detailed costing, it seems likely that the particular dense design being considered may cost 3 – 4 times as much as the sparse design for the first variant considered and be about equal in the second case (of equal receiver chain number).

In terms of performance, the dense design considered here delivers a *potential* FOV which is about 7 (or 2.5) times as large as the sparse design, although this would require an output data rate per station which is about 3 (or one) times larger. In fact, both designs vastly exceed (by about an order of magnitude) the maximum station band-width currently being considered by the SKA Data Transport working group of about 12 GHz. The station output band-width and correlator capacity may well become important constraints for the FOV which can be put to practical use. A realistic goal should be accommodation of 50 – 120 GHz (per polarization) per station out to a few 10's of km from the core. With this station data rate the dense design provides a FOV about twice that of the sparse design. The two concepts have equal sensitivity at about 500 MHz, with the relative sensitivity improving for the dense array toward 1000 MHz and the sparse array improving dramatically to lower frequencies.

For comparison we also plot the sensitivity of an array of small paraboloids which we assume to be 10,000 m$^2$/K above 1 GHz and with an aperture efficiency which roles off as $(1-e^{(-\nu/200 \text{ MHz})})$ yielding a 50% reduction in A$_{eff}$ by 150 MHz. The same model of T$_{sys}$ dependence with frequency is assumed as above.

Figure 6 nicely contrasts the frequency regimes where dishes and sparse aperture arrays excel. In the frequency interval between 700 and 300 MHz there is an almost quadratic increase of sensitivity of the sparse aperture array which translates in physical terms to an almost equal mass sensitivity to red-shifted neutral hydrogen between z = 1 and z > 3.5. Individual galaxies could still be detected in realistic integration times out to z > 5 (see Braun, 2006) with a 100% SKA. Even a 10% SKA of this sensitivity would yield samples of 10$^6$ HI galaxies in the z = 1 – 2 range in a 100 day survey. Comparable performance from an array of paraboloids in this frequency range would require almost an order of magnitude larger number of dishes than currently assumed, which is quite out of the question. The same is true of the number of dense aperture array antennas. Dense aperture arrays, as can be seen from Fig. 6, *compete with rather than complement* the performance of small dishes for ν = 0.3 – 1 GHz. Since the dishes will be essential in any case to provide performance at ν > 1 GHz, it is unclear why one would consider employing dense aperture arrays in the SKA.

While it lies outside of the scope of this document to consider the small dish component in detail, one possibility might be that a total of 4000 small dishes are distributed over 300 stations, corresponding to about 16 dishes per station. In any case, the footprint of a small dish sub-station would need to be about 100m in diameter to keep shadowing losses down to an acceptable level. One can then imagine a standard SKA "super-station" with the three basic sub-stations; the small dish sub-station extending over about 100m, together with the sparse high-band (100m) and sparse low-band sub-stations (340m). The SKA core, of 5 km diameter would be almost uniformly filled with 150 of these "super-stations", while the remaining 150 would be distributed for optimized imaging performance out to the maximum baselines. This type of homogenous distribution of "super-stations" is in marked contrast to some other suggestions that the low frequency SKA capabilities be confined to only the SKA core. There are compelling reasons to obtain wide-field sub-arcsec imaging performance all the way down to 70 MHz. Data transport costs may necessitate a lower station output band-width on 1000 km scales, but this transition from wide-field to narrow-field performance should be kept as far from the core as practical.

## 5. Some Open Issues

Several areas require further research and optimization:

- How much station band-width can actually be accommodated in the SKA (implying how much FOV can actually be utilized)?
- Are 2-D radiating antenna elements perhaps a viable, low-cost alternative to the current 3-D designs?
- What are the required qualities of station beams at sub-GHz frequencies that won't limit the SKA dynamic range?
- What are the actual properties of the sparse aperture array station beams being deployed for LOFAR?

## 6. Conclusions

Many of the anticipated scientific applications of the SKA involve some form of large area survey. For these applications it is critical to consider the "Survey Speed" $\propto$ FOV$\times S^2$ in assessing the performance of a particular SKA design. Even in the most extreme example of an ultra-wide-FOV application, all-sky transient monitoring, one can imagine achieving superior survey performance by a (suitably rapid) scanning of the sky by a more sensitive system. For the remainder of the foreseen SKA applications the FOV is essentially irrelevant, since the anticipated targets have small angular size and a previously known position. In both cases, it is clear that the primary design criterion should be sensitivity, while the FOV remains secondary.

We have assessed the performance of sparse and dense aperture arrays against this background. The distinction between sparse and dense arrays only arises when a significant operating band-width is required from a given antenna system. If we had the freedom to employ a different antenna system for each narrow band, there would be very little discussion of the optimum antenna spacing to employ. If, on the other hand, we require more realistic relative band-widths of 3:1 or more from each antenna system that is deployed, then we encounter dramatic differences in performance from the different choices of mean antenna spacing. The sensitivity penalty of closely packed antennas can be more than an order of magnitude in this case. Dense aperture arrays would then only offer an interesting option if they could be fabricated at a comparable cost per geometric area as their sparse counterparts, despite having an order of magnitude higher density of active components. The likelihood of this being realized is very small indeed.

Specific straw-man aperture array designs have been contrasted in terms of their likely cost and performance. Very attractive performance can be achieved with a sparse design that covers the 70 – 700 MHz range with two antenna systems that share receiver resources and yield a sensitivity of 10,000 – 20,000 m$^2$/K. System complexity (in terms of beam formation and number of digital receiver chains) represents only a modest increase over that being deployed for LOFAR. Here is a high performance system which can already be realized with today's (rather than tomorrow's) technology. Such an aperture array forms a cost-effective complement to the performance of small dishes which will be required in any case to provide SKA coverage for frequencies greater than 1000 MHz.

## Acknowledgements

We acknowledge very useful feed-back on earlier versions of this memo from Jaap Bregman, Ger de Bruyn, Peter Hall and Richard Schilizzi.